\documentclass[twocolumn,english,floatfix]{revtex4-2}
\usepackage[T1]{fontenc}
\usepackage{color}
\usepackage{amsmath}
\usepackage{graphicx}
\usepackage{braket}
\usepackage{bbold}
\usepackage{natbib}

\usepackage{tabularx}

\usepackage{multirow}
\usepackage{siunitx}
\usepackage[utf8]{inputenc}

{
\makeatletter
\def\frontmatter@thefootnote{%
 \altaffilletter@sw{\@fnsymbol}{\@fnsymbol}{\csname c@\@mpfn\endcsname}%
}%
\makeatother

\newcommand{\gstate}{\ket{\text{0}}}
\newcommand{\notg}{\ket{\overline{\text{0}}}}
\newcommand{\estate}{\ket{\text{1}}}
\newcommand{\fstate}{\ket{\text{2}}}
\newcommand{\tf}{\ket{\tilde{2}}}
\newcommand{\h}{\ket{\text{3}}}
\renewcommand{\eqref}[1]{\mbox{Eq.~(\ref{#1})}}

\newcommand{\figref}[1]{Fig.~\hyperref[#1]{\ref*{#1}}}
\newcommand{\figpanel}[2]{Fig.~\hyperref[#1]{\ref*{#1}(#2)}}
\newcommand{\figpanels}[3]{Fig.~\hyperref[#1]{\ref*{#1}(#2)-(#3)}}
\newcommand{\figpanelNoPrefix}[2]{\hyperref[#1]{\ref*{#1}(#2)}}

\usepackage[colorlinks=true,urlcolor=blue,citecolor=blue,linkcolor=blue]{hyperref}

\makeatother

\usepackage{babel}

\begin{document}
	
	\title{Transmon qubit readout fidelity at the threshold for quantum error correction without a quantum-limited amplifier}
	\def\andname{\hspace*{-0.5em},}
	\author{Liangyu Chen$^{1,}$}
	\email[email: ]{liangyuc@chalmers.se}
	\email[\\email: ]{jonas.bylander@chalmers.se}
	\email[\\email: ]{tancredi@chalmers.se }
	\author{Hang-Xi Li$^{1}$, Yong Lu$^{2}$, Christopher W. Warren$^{1}$, Christian J. Kri\v{z}an$^{1}$,  Sandoko Kosen$^{1}$, Marcus Rommel$^{1}$, Shahnawaz Ahmed$^{1}$, Amr Osman$^{1}$, Janka Bizn\'{a}rov\'{a}$^{1}$, Anita Fadavi Roudsari$^{1}$,  Benjamin Lienhard$^{3,4}$, Marco Caputo$^{5}$, Kestutis Grigoras$^{5}$, Leif Gr\"{o}nberg$^{5}$, Joonas Govenius$^{5}$, Anton Frisk Kockum$^{1}$, Per Delsing$^{1}$, Jonas Bylander$^{1,*}$, Giovanna Tancredi$^{1,*}$}

	\affiliation{$^{1}$Department of Microtechnology and Nanoscience, Chalmers University of Technology, 41296 Gothenburg, Sweden.\\
	$^{2}$3rd Institute of Physics, University of Stuttgart, 70569 Stuttgart, Germany.\\
	$^{3}$Department of Chemistry, Princeton University, Princeton, NJ 08544, USA\\
	$^{4}$Department of Electrical and Computer Engineering, Princeton University, Princeton, NJ 08544, USA\\
	$^{5}$VTT Technical Research Centre of Finland, FI-02044 VTT, Finland}
	
	\begin{abstract}
	
    
    High-fidelity and rapid readout of a qubit state is key to quantum computing and communication, and it is a prerequisite for quantum error correction. We present a readout scheme for superconducting qubits that combines two microwave techniques: applying a shelving technique to the qubit that effectively increases the energy-relaxation time, and a two-tone excitation of the readout resonator to distinguish among qubit populations in higher energy levels. Using a machine-learning algorithm to post-process the two-tone measurement results further improves the qubit-state assignment fidelity. We perform single-shot frequency-multiplexed qubit readout, with a \SI{140}{ns} readout time, and demonstrate 99.5\,\% assignment fidelity for two-state readout and 96.9\,\% for three-state readout—without using a quantum-limited amplifier.
    
		
	\end{abstract}
	
	\date{\today}
	
	\maketitle
	
\section{Introduction}

With the recent demonstrations of quantum error correction \cite{Google2021, krinner2021realizing,zhao2022realization,Google2022}, superconducting circuits are one of the leading platforms towards the realization of a fault-tolerant quantum computer \cite{Preskill1997,Fowler2012, Martinis2015}. However, despite the remarkable progress, achieving fast and high-fidelity single-shot readout of the qubits' states remains a challenge. As a comparison,  while the two-qubit-gate fidelities are approaching the $0.1\,\%$ error threshold \cite{Foxen2020,Yuan2020,Negrneac2021,Sung2021}, readout errors are typically at the $1\,\%$ level for two-state readout \cite{Walter2017,Heinsoo2018,Elder2020, Sunada2022}. Similarly, the implementation of high-fidelity single- and two-qubit gates takes between 10 and 100~ns \cite{Negrneac2021,Sung2021,Blais2021}, while a readout measurement can take from hundreds of nanoseconds to a few microseconds \cite{Walter2017,Heinsoo2018, Elder2020,Blais2021}. Further improvement in the readout of superconducting qubits is therefore crucial to reliably cross the threshold of efficient error correction, which is estimated to be less than $0.5\,\%$ for the break-even point \cite{Martinis2015}. Moreover, having a fast and high-fidelity measurement scheme can boost the repetition rate for both quantum-computing and quantum-communication applications \cite{Campagne2018,Kurpiers2018,Kurpiers2019,Ilves2020,Magnard2020} and is essential for achieving fast reset protocols \cite{Dassonneville2020,Sunada2022}.\par

In superconducting circuits, the state of a superconducting qubit is generally read out by detecting the dispersive frequency shift of a resonator coupled to the qubit \cite{Blais2004}. The predominant source of error is the relaxation of the qubit from the excited ($\estate$) to the ground ($\gstate$) state during the readout. On short time scales, this error grows almost linearly with the ratio between the readout time $\tau_{r}$ and the qubit relaxation time $T_{1}$ \cite{Walter2017}, and can be mitigated by reducing $\tau_{r}$. Various high-power readout schemes have been exploited to decrease the measurement time  \cite{Reed201010,Gusenkova2021}. Furthermore, Purcell filters \cite{Mallet2009,Reed2010,Jeffrey2014,Sete2015} and quantum-limited or near-quantum-limited amplifiers \cite{Yurke1989,Mutus2014,Macklin2015,Renger2021} have been implemented and, with the combination of both, a readout fidelity exceeding $99\,\%$ within $100~\rm{ns}$ has been demonstrated \cite{Walter2017,Heinsoo2018}.\par

Here we report the implementation of a novel readout scheme that boosts the state-assignment fidelity and increases the effective qubit relaxation time during readout. Our readout strategy exploits the higher energy levels of the qubit \cite{Martinis2002,Mallet2009,Peterer2015,Jurcevic2020,Elder2020,Wang2021} and introduced a two-tone probing of the resonator to enhance the readout fidelity of multiple states. We demonstrate single-shot readout fidelity up to $99.5\,\%$ ($96.9\,\%$) for two-state (three-state) discrimination within $140~\rm{ns}$ without using a quantum-limited amplifier. The techniques we present here offer significant protection against decay during readout, are straightforward to implement, and can be readily integrated in state-of-the-art quantum-computing devices. \par




\section{RESULTS}
\subsection{Experimental setup}
\label{sec_setup}

\begin{figure}[ht]
\centering
 \includegraphics[width=2.5in, keepaspectratio=true]{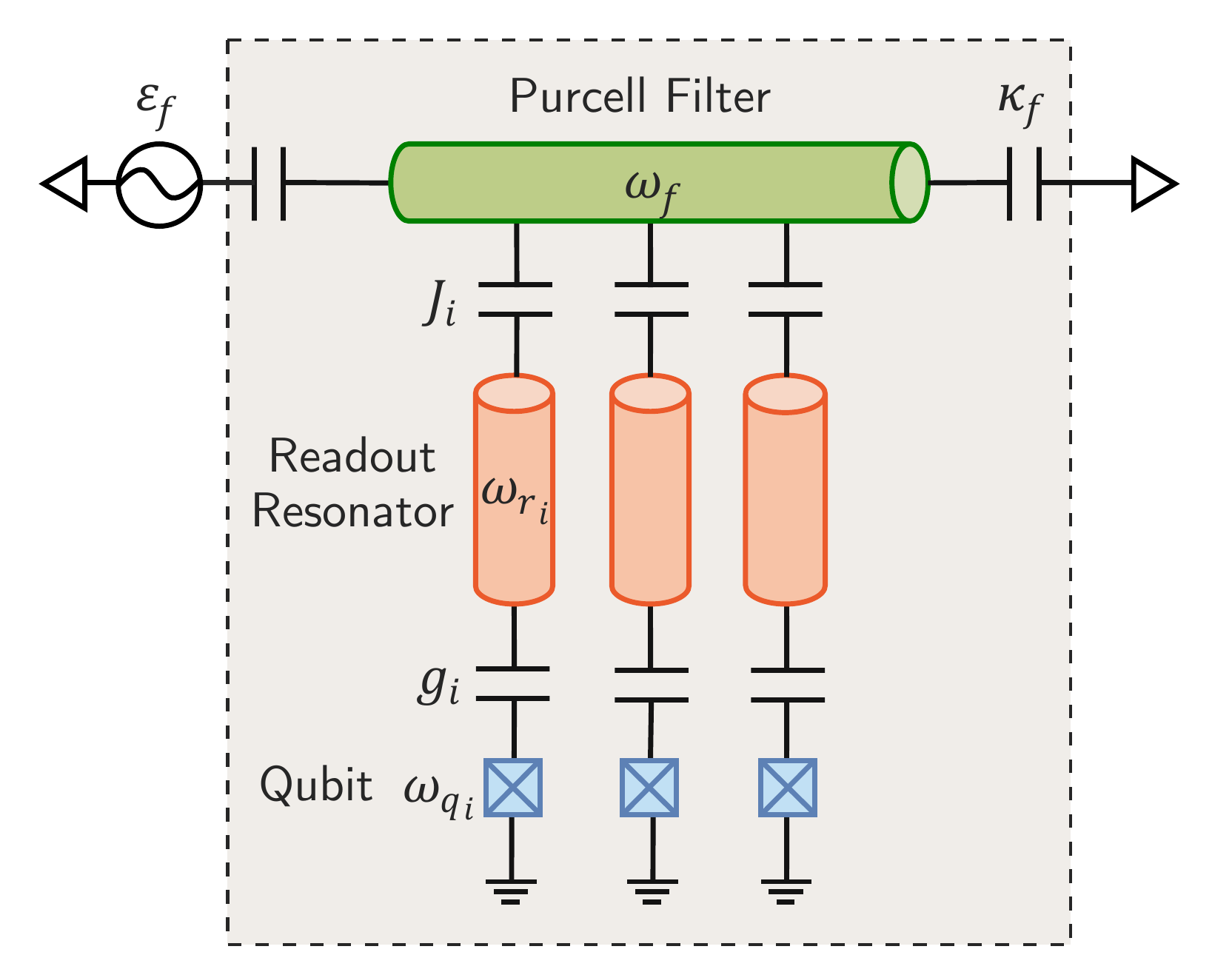}
 \caption {\textbf{Circuit schematic of the common-mode Purcell filter.} The Purcell filter is a $\lambda/2$ coplanar waveguide resonator centered at $\omega_{f}$, defined by a capacitor on each side. The filter is embedded in the readout feedline, and driven by the field $\epsilon_{f}$. The output capacitance, represented by the Purcell-filter linewidth $\kappa_{f}$, is around an order of magnitude larger than the input capacitance such that the signal is guided towards the output port to measure transmission. Multiple resonators of resonant frequeny $\omega_{r_{i}}$ couple to the Purcell filter with strength $J_{i}$ within the filter bandwidth. The individual resonators are capacitively coupled with strength $g_{i}$ to the qubits with transition frequency $\omega_{q_{i}}$.}
\label{fig_resonator}
\end{figure}
\noindent

The device design and fabrication is described in Ref. \cite{Kosen2022} with the circuit schematic shown in \figref{fig_resonator}. The device consists of three fixed-frequency transmon qubits \cite{Koch2007} with transition frequencies $\omega_{q_{i}}/2\pi$ at $5.36$, $5.40$, and $5.46$~GHz for $i = 1$, $2$, $3$, respectively. Each qubit is coupled with a strength $g_{i}$ to a readout resonator of frequency $\omega_{r_{i}}/2\pi = 6.45$, $6.61$, and $6.74$~GHz. The three resonators are coupled with a strength $J_{i}$ to a common Purcell filter that is embedded in the readout feedline \cite{Jeffrey2014}. The Purcell-to-resonator coupling rates $J_{i}/2\pi$ are designed to be at 60~MHz, while the qubit-resonator coupling rates $g_{i}/2\pi$ are much larger, about 250~MHz. The Purcell filter is centered at $\omega_{f}/2\pi = 6.726$~GHz with a linewidth of $\kappa_{f}/2\pi = 820.9$~MHz. The theoretical analysis and design details are discussed in the Supplementary Information~I. The device is cooled down to 10~mK and a microwave setup is used to measure the signal transmitted through the feedline. The complete experimental setup and device parameters resulting from basic characterization are reported in the Supplementary Information~II.\par


\begin{figure}[ht]
\centering
 \includegraphics[width=3.4in, keepaspectratio=true]{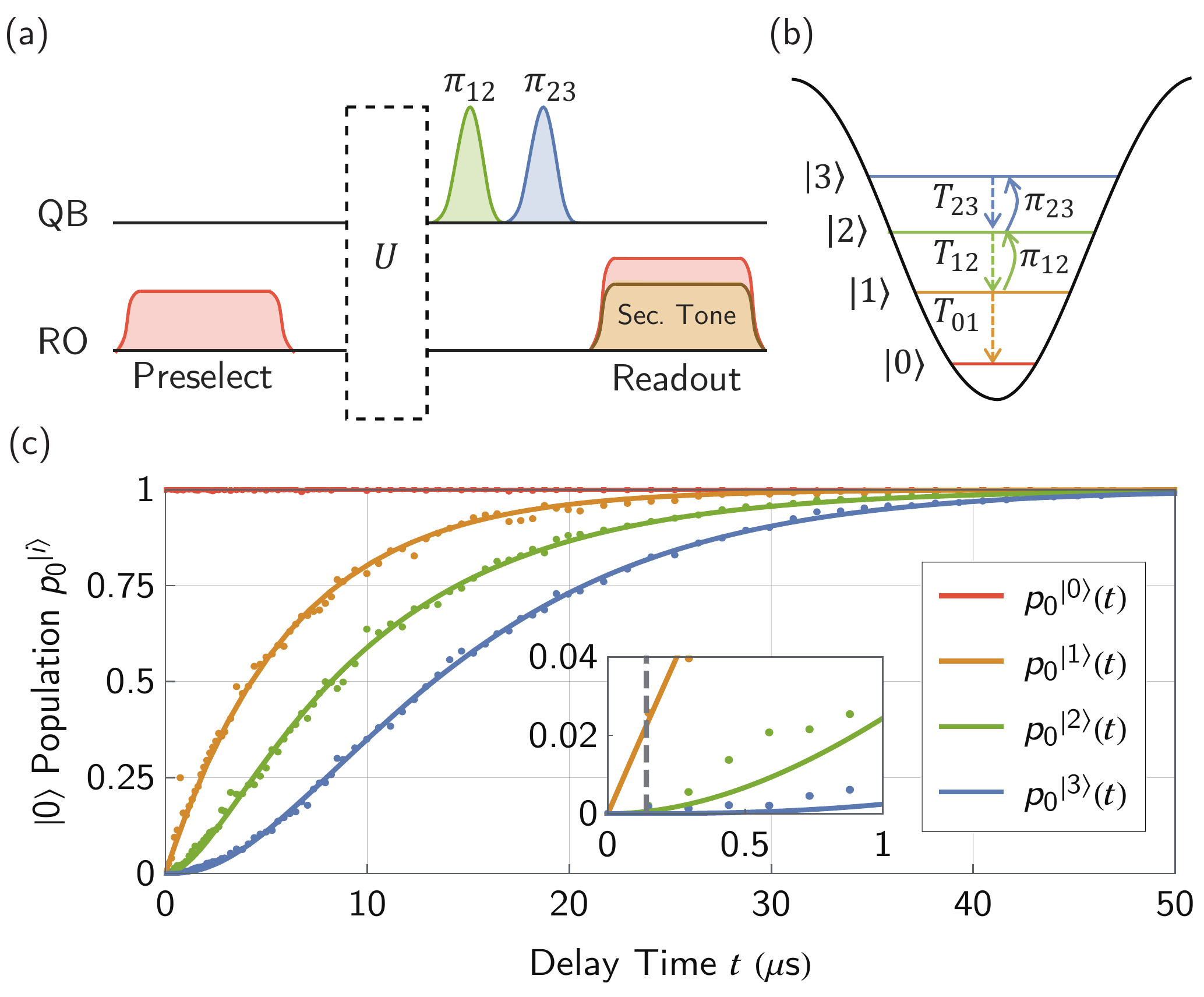}
 \caption {\textbf{Schematics and implementation of the shelving and two-tone readout technique.} \textbf{(a)} A $\pi_{12}$ and a consecutive $\pi_{23}$ pulse are inserted between any experimental sequence $\textit{\textbf{U}}$ and the final readout pulse consists of two readout tones. This scheme can be implemented in any multi-level quantum processor platform without modifying the hardware design. \textbf{(b)} Qubit population can be transferred to the desired energy level with consecutive $\pi_{ij}$ pulses. The population in state $\ket{j}$ decays to $\ket{i}$ with a rate $1/T_{ij}$. \textbf{(c)} The ground state $\gstate$ population $p_{0}$ is plotted as a function of the delay time $t$ after the transmon is initially prepared in $\gstate$, $\estate$, $\fstate$, or $\h$. Points represent experimental data for Qubit 2 while continuous lines show fits of the data according to \eqref{eqn_RateEfPulse}. The inset shows the population at short time scales with the dashed line marking the duration  $\tau_r = 140$\,ns of the readout pulse.
 }
\label{fig_T1efh}
\end{figure}
\noindent

\subsection{Exploiting higher energy levels}
\label{sec_higher_energy}

Typically, the qubit-excited-state decay is a major error source during readout, whose minimization requires performing the measurements in the shortest possible time \cite{Walter2017,Heinsoo2018}. To further mitigate this error, we first implement a shelving scheme that exploits the higher energy levels \cite{Martinis2002,Mallet2009,Peterer2015,Jurcevic2020,Elder2020,Nathanael2021,Wang2021}. The pulse scheme is shown in \figpanel{fig_T1efh}{a}; a $\pi_{12}$ and a $\pi_{23}$ pulse are applied consecutively prior to the readout pulse so that the qubit population originally in the $\estate$ state is transferred to the $\h$ state before the readout. Thus, the qubit population that was in $\estate$ will take a longer time to decay to $\gstate$ as the main relaxation channel is through cascading single-photon emission down the energy ladder, illustrated in \figpanel{fig_T1efh}{b}.\par

To quantify the possible improvement, we measure the population of the ground state $p_{0}(t)$ as a function of time $t$ when the qubit is prepared in $\gstate$, $\estate$, $\fstate$, and $\h$. The data for Qubit 2 is shown in \figpanel{fig_T1efh}{c}. State preparation and measurement (SPAM) errors are mitigated by applying the inverse of the assignment matrix to the measurement results. Then, the most probable physical state is acquired with a maximum likelihood estimator \cite{Baumgratz2013}. We find that when the qubit is prepared in $\estate$, the $\estate$-state population decays during the readout by $\epsilon=1-e^{-\tau_r/T_{01}}=2.24\,\%$, with relaxation time $T_{01}=\SI{6.18}{\micro\second}$, giving a significant contribution to the readout error. The duration of the readout pulse $\tau_r=\SI{140}{ns}$ is minimized by optimizing the readout-pulse amplitude without introducing significant readout-induced mixing that deteriorates readout performance \cite{Schuster2005,Gambetta2006,Schuster2007}.
\par
We calculate the population $p_{i}(t)$ in the $\ket{i}$ state with the following rate equations:
\begin{equation}
\label{eqn_RateEfPulse}
\begin{split}
\dot{p_{0}}(t) &= p_{1}(t)/T_{01},\\
\dot{p_{1}}(t) &= p_{2}(t)/T_{12} - p_{1}(t)/T_{01},\\
\dot{p_{2}}(t) &= p_{3}(t)/T_{23}-p_{2}(t)/T_{12},\\
\dot{p_{3}}(t) &= -p_{3}(t)/T_{23},
\end{split}
\end{equation}
 \noindent
where $T_{ij}$ is the relaxation rate from the $\ket{j}$ to the $\ket{i}$ state as illustrated in the level diagram of \figpanel{fig_T1efh}{b}. The anharmonicity of the transmon is sufficient such that the nonsequential decay through multilevel channels is exponentially suppressed. The contribution of direct decay from $\fstate$ to $\gstate$ is found to be two orders of magnitude smaller than that from $\fstate$ to $\estate$ \cite{Peterer2015} and is neglected in the equations. We can solve for the evolution of any $\ket{i}$-state population as a function of time $t$ when the qubit is initialized in the $\ket{j}$ state, denoted as $p^{\ket{j}}_{i}(t)$. Specifically, we first solve for $p^{\fstate}_{0}(t)$ and assume $p^{\fstate}_{2}(0) = 1$ in the absence of any error and neglect the effect of higher-energy levels by using the initial conditions $p^{\fstate}_{0} = p^{\fstate}_{1} = p^{\fstate}_{3} = 0$. We find:
\begin{equation}
\label{eqn_p0EffhPulse}
\begin{split}
    p^{\fstate}_{0}(t) &= 1-
    \frac{T_{01} \; e^{-t/T_{01}}}{T_{01}-T_{12}}+\frac{T_{12} \; e^{-t/T_{12}}}{T_{01}-T_{12}}.
\end{split}
\end{equation}
 \noindent
where $T_{12}$ is approximately a factor of $\sqrt{2}$ smaller than $T_{01}$ for typical transmon parameters. The second and third terms in \eqref{eqn_p0EffhPulse} are two decaying functions with opposite signs, hence the net result is no longer purely exponential. We also solve the rate equations of the system in \eqref{eqn_RateEfPulse} for $p^{\h}_{0}$, when the qubit is prepared in $\h$, with the initial conditions $p^{\h}_{0}= p^{\h}_{1} = p^{\h}_{2} = 0$, and find the following:
\begin{equation}
\label{eqn_p0EffhPulse23}
\begin{split}
    p^{\h}_{0}(t) = & \: 1-\frac{T_{01}{}^2 \;e^{-t/T_{01}}}{(T_{01}-T_{12}) (T_{01}-T_{23})} \\
      & +\frac{T_{12}{}^2 \;e^{-t/T_{12}}}{(T_{01}-T_{12}) (T_{12}-T_{23})}\\
     & -\frac{T_{23}{}^2 \;e^{-t/T_{23}}}{(T_{01}-T_{23}) (T_{12}-T_{23})}.
\end{split}
\end{equation}


This equation contains a combination of exponential decays with different signs as well. The solutions are used to fit the data in \figpanel{fig_T1efh}{a}, and we find excellent agreement between the data and the model.\par
For Qubit 2 in particular, we find that the readout error from the $\estate$ state is reduced from $2.24\,\%$ to $0.057\,\%$ after the application of a $\pi_{12}$ pulse, and to $8.7\times10^{-4}\,\%$ after a $\pi_{23}$ pulse. This is calculated by taking the difference between $p^{\ket{2}}_0(t=\tau_r)$ and $p^{\ket{2}}_0(0)$ in \figpanel{fig_T1efh}{a}. The remaining  error with the addition of a $\pi_{12}$ pulse is equivalent to the decay error of a qubit with $T_{01}=245~\mu$s using the standard readout method. Thus, here we achieve larger than an order-of-magnitude improvement in effective $T_{01}$ during the readout. For a longer-lived qubit, the percentage of decay errors that can be suppressed with shelving continues to grow closer to unity \cite{Wang2021}. However, other error contributions will likely become more prominent at this level.\par

\subsection{Two-state readout with a primary tone}

\begin{figure}[t]
\centering
 \includegraphics[width=3.4in, keepaspectratio=true]{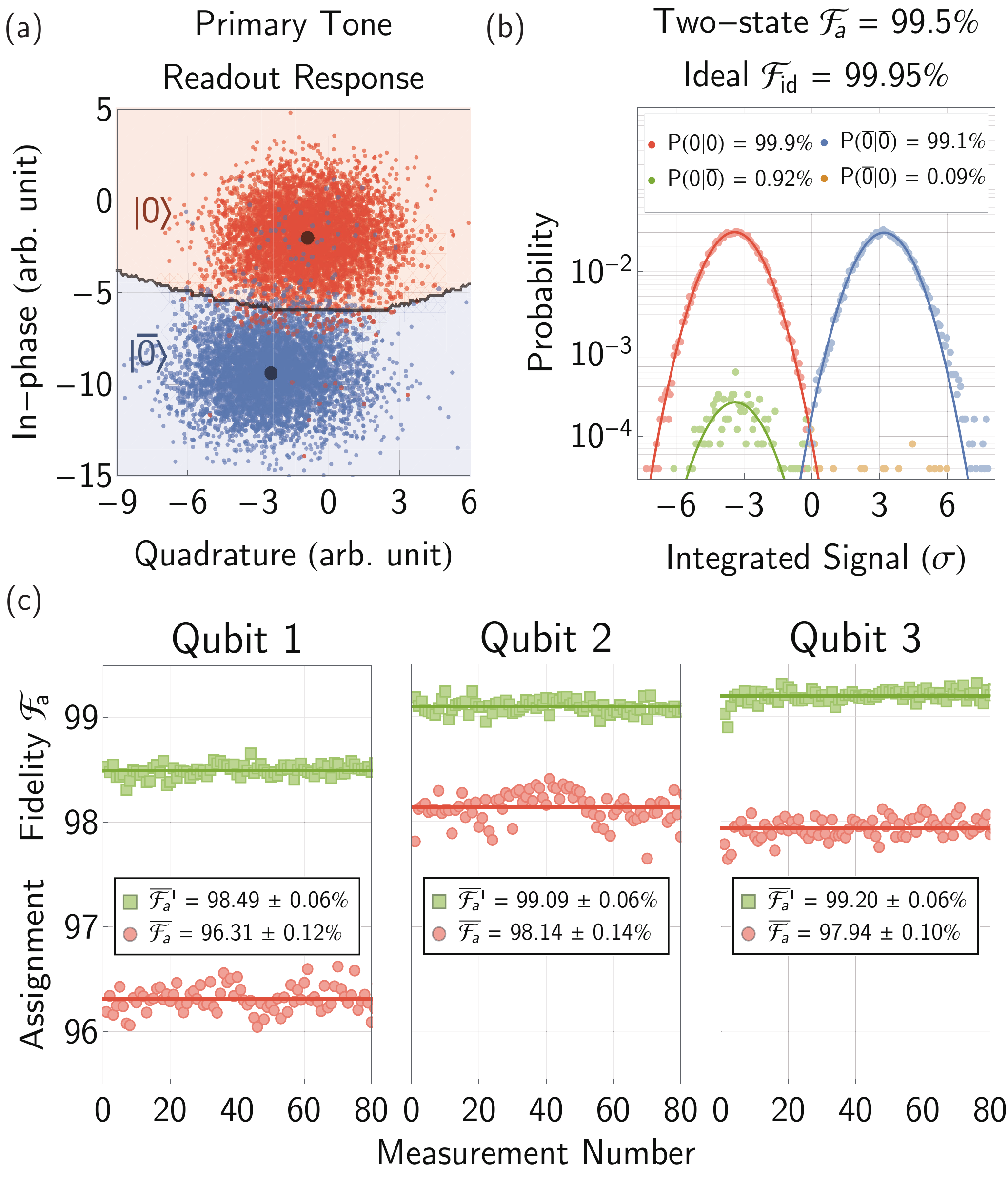}
 \caption{\textbf{Single-shot readout results of discriminating between $\gstate$ and $\notg$.} \textbf{(a)} Integrated readout signal in the IQ plane with Qubit 2 prepared in $\gstate$ and $\estate$. With the consecutive $\pi_{12}$ and $\pi_{23}$ pulses implemented prior to a \SI{140}{ns} readout, we distinguish between $\gstate$ and $\notg$ (NOT $\gstate$). \textbf{(b)} The IQ-plane signals in (a) are projected onto an optimal axis, and the resulting histogram is fitted with a Gaussian distribution. The horizontal axis is normalized by the standard deviation $\sigma$. The calculated assignment fidelity $\mathcal{F}_{a}$ and ideal fidelity $\mathcal{F}_{id}$ are shown above the plot. The conditional probabilities $P(i\lvert j)$ represent the probabilities of measuring state $\ket{i}$ given that the qubit is prepared in state $\ket{j}$. \textbf{(c)} Simultaneously measured single-shot readout assignment fidelities for the three qubits with ($\overline{\mathcal{F}_{a}'}$) and without ($\overline{\mathcal{F}_{a}}$) the application of the $\pi_{12}$ and $\pi_{23}$ pulses. The error statistics are calculated from the standard deviation of the measured set.}
    \label{fig_ssroQubit}
\end{figure}
\noindent


Having optimized the state-preparation pulses (see Methods section), we fine-tune the readout frequency to maximally separate the 2D in-phase and quadrature (IQ) histograms corresponding to the $\gstate$ and $\estate$ states, as shown in \figpanel{fig_ssroQubit}{a}. Higher energy levels are indistinguishable from $\estate$ in this configuration, and we can only distinguish between $\gstate$ and $\notg$ (NOT $\gstate$). To calibrate the readout, we prepare the qubit in either $\gstate$ or $\estate$ and add the $\pi_{12}$ and $\pi_{23}$ pulses to transfer the $\estate$-state population to the $\h$ state before the readout, as illustrated in \figpanel{fig_T1efh}{a}. First, we start with a single readout tone in the pulse which is referred to as the primary tone. To discard the results for which the initial state is not $\gstate$, we include a preselection pulse, i.e., an additional readout measurement before the sequence starts \cite{Walter2017, Rist2012}. Through this preselection procedure, thermal and residual populations in the qubits are filtered from the outcomes.\par

We perform simultaneous readout of all three qubits and calculate the single-qubit readout assignment fidelity $\mathcal{F}_a = 1-[P(0\lvert \overline{0})+P(\overline{0}\lvert 0)]/2$, where $P(i\lvert j)$ is the classification probability, i.e., the probability that the $\ket{i}$ state is assigned given that the $\ket{j}$ state is initially prepared. This measure of readout fidelity takes all the error contributions into account, including imperfect state preparation before the readout sequence. The assignment fidelity for 80 repetitions, each containing 50,000 shots, is shown in \figpanel{fig_ssroQubit}{c} with ($\overline{\mathcal{F}_a'}$) and without ($\overline{\mathcal{F}_a}$) implementation of the shelving technique. The data demonstrates a reduction in the overall error rate by 57$\,\%$ on average for the three qubits with the introduced readout scheme. We also compute the ideal readout fidelity $\mathcal{F}_{id}$ by integrating the overlapping area of the Gaussian probability distributions after projecting 
the IQ data onto an optimal axis \cite{Magesan2015}:
\begin{equation}
\label{eqn_fid}
\mathcal{F}_{id} =  \frac{1}{2}\left[1+\mathrm{erf}\left(\sqrt{\frac{SNR^{2}}{8}}\right)\right],
\end{equation}
\noindent
where the signal-to-noise ratio is
\begin{equation}
\label{eqn_p1EffhPulse}
 SNR = \frac{\left\lvert  \left\langle S_{0} \right\rangle-\left\langle S_{\overline{0}} \right\rangle \right\lvert }{\sigma_{S_{0}}},
\end{equation}
\noindent
with $S_{i}$ being the set of measurement outcomes and $\sigma^2_{S_{0}}$ being the variance of the data set. For Qubit 2, the best assignment fidelity is $99.5\,\%$ while the ideal fidelity exceeds $99.95\%$; see \figpanel{fig_ssroQubit}{b} for detailed histograms. The error $\epsilon$ from qubit decay ideally no longer contributes to the remaining error, $P(0\lvert \overline{0})=0.92\,\%$, when measuring the excited state. This suggests that the fidelity is predominantly limited by other sources of error, such as state preparation \cite{Walter2017} and measurement-induced mixing \cite{Schuster2005,Gambetta2006,Schuster2007,Sank2016,Slichter2012}.\par





\subsection{Three-state readout with two tones}


Choosing the optimal readout frequency to attain the best two-state assignment fidelity leaves other higher-energy states indistinguishable from each other. However, the information of the $\fstate$-state population is crucial to detect leakage errors during gate calibrations and algorithms \cite{krinner2021realizing}. To access this information, we introduce a secondary readout tone with the frequency that maximizes the separation between $\estate$ and $\fstate$. This pulse is multiplexed with the primary pulse to perform the readout measurements simultaneously.\par

\begin{figure}[t!]
\centering
 \includegraphics[width=3.4in, keepaspectratio=true]{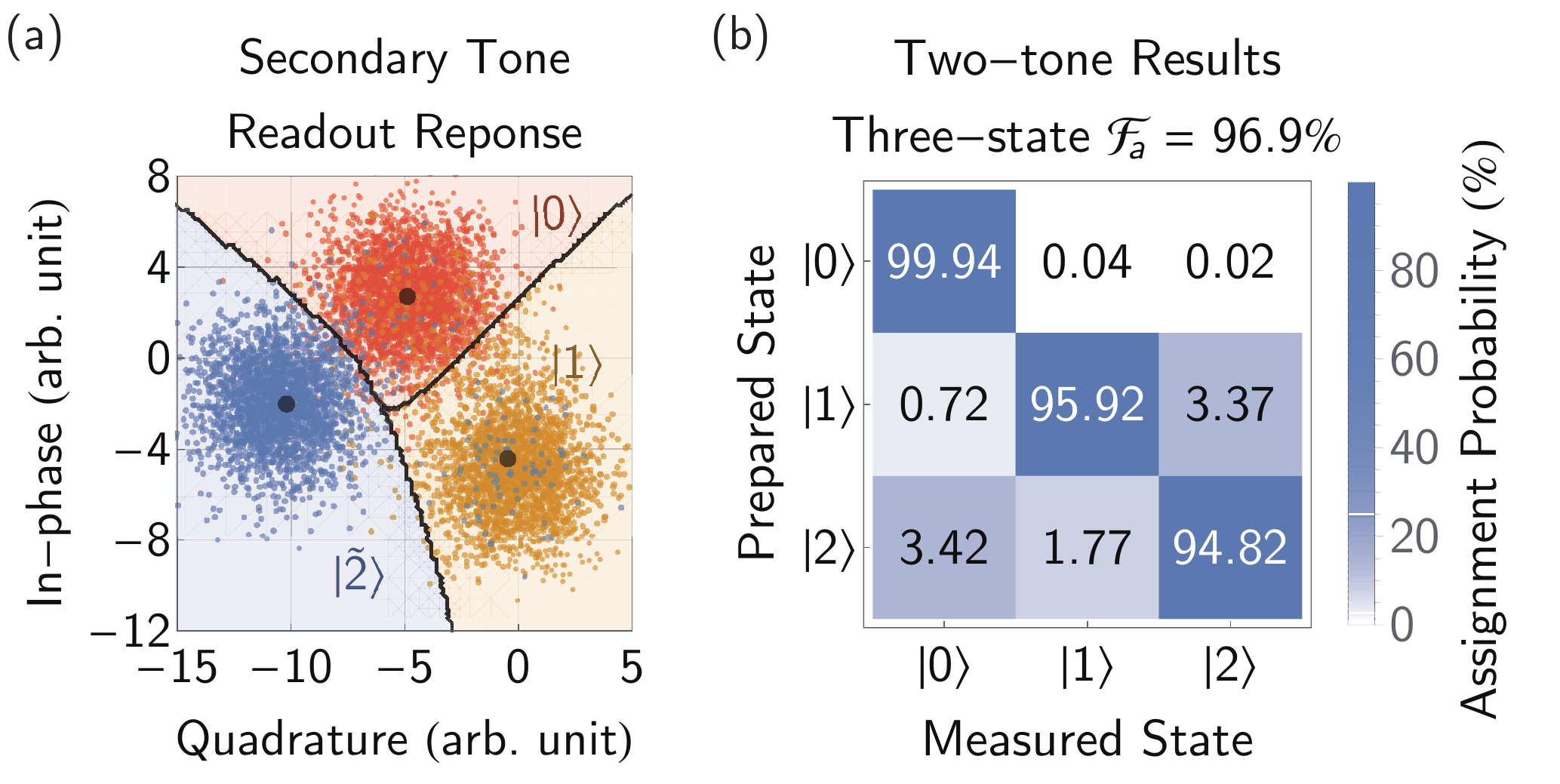}
 \caption{\textbf{Single-shot readout results for discriminating between the $\gstate$, $\estate$, and $\fstate$ state.} \textbf{(a)} Integrated single-shot readout signal of the secondary tone for Qubit 2. The $\gstate$ and $\estate$ states are distinguishable from the rest, while $\fstate$ and $\h$ have a significant overlap, and are therefore being combined into a single classification: $\tf$. The frequency of the primary tone is identical to that shown in \figpanel{fig_ssroQubit}, which maximizes the distinction between $\gstate$ and $\notg$. The dots indicate the centers of the Gaussian distributions. \textbf{(b)} Three-state assignment matrix with the two-tone readout for Qubit 2, reconstructed using a neural network. Note that the most significant error contributions in the two-tone readout are the misclassification between $\fstate$ and $\gstate$ as well as that  between $\estate$ and $\fstate$.}
    \label{fig_ttPulseIQ}
\end{figure}
\noindent

We also use the $\pi_{12}$ and $\pi_{23}$ pulses to implement the shelving scheme. As the initial $\estate$-level population is transferred to the $\h$ state and the $\fstate$-population is transferred to the $\estate$ state, an error in $\estate$-state assignment will occur if a cascade of decays happens from $\h$ to $\estate$. The effective relaxation time during readout is enhanced, leading to similar improvement as that discussed in Sec.~\ref{sec_higher_energy}. To quantify the improvement, we need to solve \eqref{eqn_RateEfPulse} for the evolution of population $p^{\h}_{1}(t)$. The analytical solution is
\begin{equation}
\label{eqn_p1EffhPulse23}
\begin{split}
    p^{\h}_{1}(t) =\: & \frac{T_{01}{}^2 \: e^{-t/T_{01}}} {(T_{01}-T_{12}) (T_{01}-T_{23})} \\
      &-\frac{T_{01}\:T_{12} \: e^{-t/T_{12}}} {(T_{01}-T_{23}) (T_{12}-T_{23})}\\
      &+\frac{T_{01}\:T_{23} \: e^{-t/T_{23}}} {(T_{01}-T_{12})(T_{12}-T_{23})}.
\end{split}
\end{equation}
\noindent
With the qubit being in the $\h$ state ($p^{\h}_{3}(0)=1$), the effective population accumulation in $\estate$ after a readout time of $\tau_{r} = 140$~ns is $0.035\,\%$. This is two orders of magnitude smaller than the error from a direct decay of the $\fstate$-state population with rate $1/T_{12}$, corresponding to $2.65\,\%$. Therefore, the contribution from energy relaxation to the three-state readout error should be reduced by a similar factor.\par

The secondary tone is tuned up in the presence of the primary pulse. An initial estimate for the secondary readout frequency is where the $\estate$- and $\fstate$-state responses are maximally separated in the IQ-plane (see Methods section). We then fine-tune this frequency such that $\gstate$ and $\h$ are distinguishable from each other as well. After optimization, the two readout pulses are typically a few MHz apart and are multiplexed in a single waveform for the readout. The transmitted signal is then processed with standard multiplexed readout techniques \cite{Heinsoo2018}. We obtain two complex voltages after signal integration, each containing a pair of in-phase and quadrature values. Overlap errors are then filtered by comparing the results in post-processing. The response of the secondary tone when Qubit 2 is prepared in $\gstate$, $\estate$, and $\fstate$, with 50,000 repetitions per state, is shown in \figpanel{fig_ttPulseIQ}{a}. Since $\fstate$ and $\h$ are indistinguishable, we combine the results together and relabel them as the $\tf$-state.\par

\begin{table}[t]
\centering
\begin{tabularx}{0.4\textwidth} { 
  | >{\centering\arraybackslash}X 
  | >{\centering\arraybackslash}X
  | >{\centering\arraybackslash}X
  | >{\centering\arraybackslash}X | }
\hline
\begin{tabular}[c]{@{}c@{}}Primary\\Result\end{tabular} & \begin{tabular}[c]{@{}c@{}}Secondary\\Result\end{tabular} & \begin{tabular}[c]{@{}c@{}}Before\\Readout\end{tabular} & \begin{tabular}[c]{@{}c@{}}Initial\\State\end{tabular}  \\ 
\hline
 $\gstate$ & $\gstate$ & $\gstate$ & $\gstate$  \\ 
 $\notg$ & $\estate$ & $\h$ & $\estate$\\
 $\notg$ & $\tf$ & $\estate$ & $\fstate$ \\
\hline
 $\gstate$ & $\notg$ & \multicolumn{2}{c|}{\multirow{2}{*}{Overlap Error}}   \\
 $\notg$ & $\gstate$ & \multicolumn{2}{c|}{}   \\
\hline
\end{tabularx}
\caption{\textbf{Truth table of the selection rule for the two-tone readout.} The first two columns are the discrimination results from the two readout tones, respectively. There is a unique initial state if the results agree with each other. Otherwise, the shots where the two readout results disagree will be counted towards overlap error and discarded.}
\label{table_tt}
\end{table}
We then formulate two methods to combine the results from the primary and secondary readout pulses to reconstruct the population initially prepared in $\gstate$, $\estate$, and $\fstate$. The first method is a truth table (see Table \ref{table_tt}) that takes the individual measurement results from the two tones as a pair of input values. There exists a unique initial state that is compatible with both results. For example, if the primary result is $\notg$ and the secondary result is $\estate$, then the initially prepared state  must be $\fstate$. In case these two results cannot reach a common decision due to overlap error, the measurement is discarded.\par

The second method to discriminate the qubit state utilizes a feedforward neural network (FNN). The network was organically developed for multiplexed readout \cite{Lienhard2022} and is adapted here to treat the two data sets as a single system (see Methods section). The input to the neural network combines the in-phase ($I[n]$) and quadrature ($Q[n]$) data from the $n^{\mathrm{th}}$ integrated signal of the two tones into a four-element vector $\{I_{1}[n],\:Q_{1}[n],\:I_{2}[n],\:Q_{2}[n]\}$. After being trained with a calibration data set, the neural network is able to classify two-tone results and give the initial qubit state as the output.

The resulting assignment matrix, shown in \figpanel{fig_ttPulseIQ}{b}, demonstrates an assignment fidelity of $96.9\,\%$ for the three-state readout. This result shows a significant improvement over the average $94.2\,\%$ assignment fidelity that we find using only a single readout pulse at an optimal readout frequency to distinguish between $\gstate$, $\estate$, and $\fstate$, with the overall error rate reduced by 47$\,\%$. The amount of suppression is achieved with the neural network that consistently outperforms the truth table by 13$\,\%$ in overall error rate reduction. \par

A feature of the resulting assignment matrix is that the population originally in $\fstate$ has a higher probability to be misidentified as $\gstate$ than as $\estate$, which is due to the shelving technique. Since the initial $\fstate$-state population is transferred to $\estate$ before measurement, decaying to the ground state is more likely than the excitation back to higher energy levels.\par





\begin{figure}[t!]
\centering
 \includegraphics [width=3.06in, keepaspectratio=true] {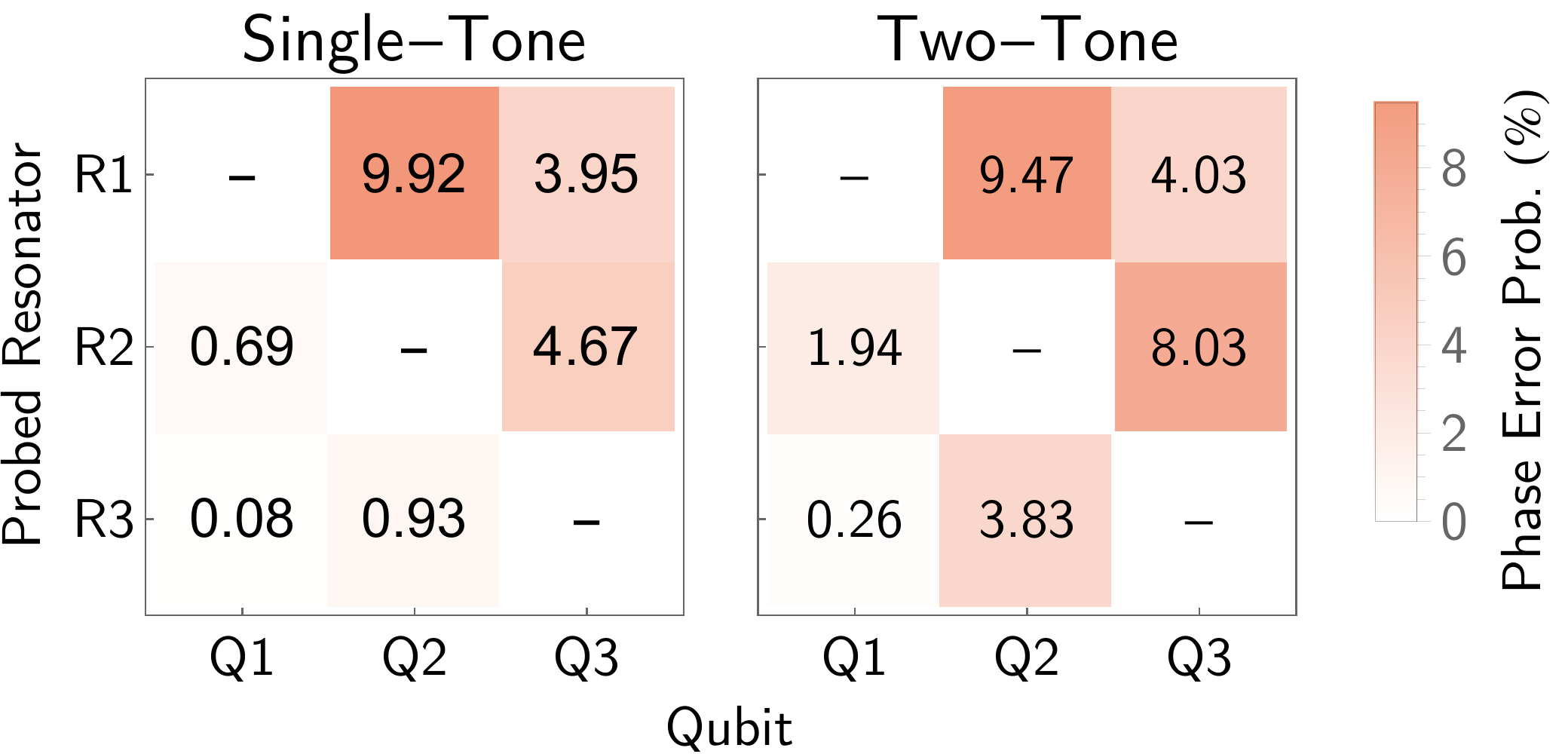}
\caption{\textbf{Measurement-induced dephasing matrix for single-tone and two-tone readout.} Each element represents the qubit dephasing while a pulse is targeting one of the readout resonators (R1, R2, or R3). Note that for the two-tone case, resonator 1 was not driven by an additional tone, so that it served as a benchmark measurement.}
\label{fig_mid}
\end{figure}
\noindent

We also investigate the effect of increased photon population in the resonators due to the secondary tone. A large photon number leads to significant measurement-induced mixing and readout crosstalk that contribute to the overall readout error \cite{Heinsoo2018, Bultink2018}. To minimize readout errors due to measurement-induced mixing, we optimize the readout amplitude such that the critical photon number $n_{\mathrm{crit}}=\Delta^{2} / 4 g^{2}$ is not exceeded with the addition of the secondary tone. To investigate readout crosstalk, we determine the measurement-induced dephasing with and without the secondary tone \cite{Heinsoo2018}. We find that the probabilities of a phase error in untargeted qubits are a factor of three larger on average due to the increase in photon number, as shown in \figref{fig_mid}. Mitigation strategies may be required if this contribution becomes significant for error-correction algorithms.\par
In the design of future devices, the qubits could be grouped into physically separated readout lines depending on their designation, e.g., ancilla or data qubits, and whether their measurements occur simultaneously. Moreover, the induced crosstalk could be further mitigated with other techniques such as machine-learning algorithms for discrimination and readout pulse shaping \cite{Duan2021,Lienhard2021,Lienhard2022}. \par
	
\section{DISCUSSION}

In conclusion, we have demonstrated that exploiting the higher energy levels of the qubit together with the implementation of a secondary readout tone lead to an improved readout-assignment fidelity of $99.5\,\%$ ($96.9\,\%$) for two-state (three-state) discrimination within 140~ns of readout time, reducing the overall error rate by $57\,\%$ ($47\,\%$) compared to our baseline.
This result could be further improved with the use of a quantum-limited parametric amplifier. The proposed pulse scheme is straightforward to implement in any measurement sequence to improve readout fidelity. \par


To further develop these readout techniques, more complex methods in the construction of the primary and secondary readout tones could be explored. More sophisticated deep neural-network methods could also be employed to aid state classification of the two-tone readout results \cite{Lienhard2022}. The possibility to generalize these techniques to further boost fidelity for multiplexed readout is a promising prospect for the future investigation of quantum computing with superconducting qubits.\par


	
\section{METHODS}
\subsection{Pulse calibration}

We optimize the parameters of the $\pi_{12}$ and $\pi_{23}$ pulses similar to the standard method developed for the $\pi_{01}$ pulse. The pulse lengths are set to be \SI{50}{ns} as a starting point. We first prepare the qubit in $\estate$ and conduct Rabi and Ramsey-like experiments between the higher energy levels to optimize the amplitude and frequency of the respective drive pulse. For shorter pulse lengths a proper derivative removal by adiabatic gate (DRAG) \cite{Motzoi2009,Gambetta2011} needs to be calibrated for the $\pi_{12}$ and $\pi_{23}$ pulses as well.\par
\begin{figure}[ht]
\centering
 \includegraphics[width=3.in, keepaspectratio=true]{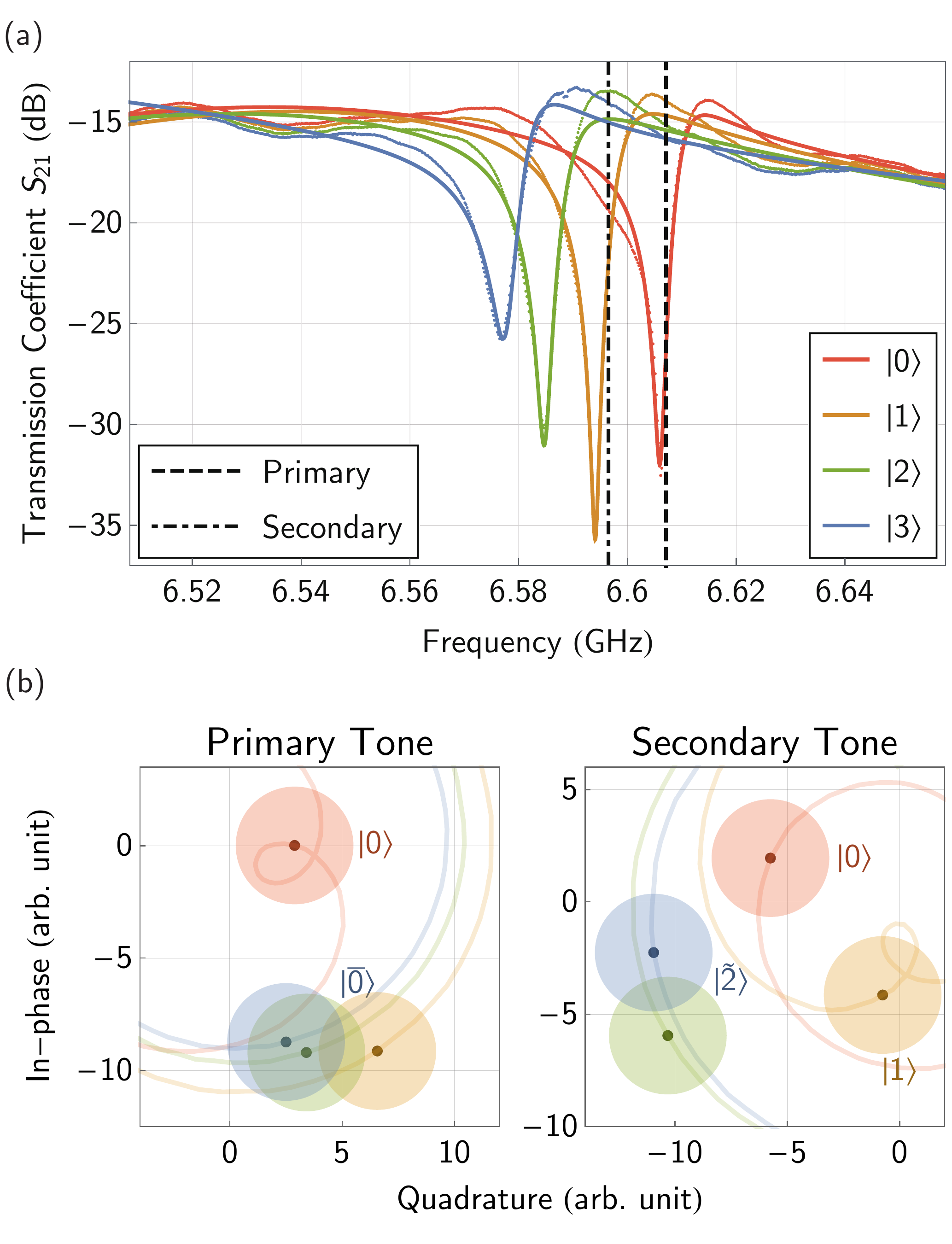}
 \caption{\textbf{Transmission coefficient $S_{21}$ as a function of the driving frequency $\omega_{d}$ centered around Resonator 2.} \textbf{(a)} Qubit-state-dependent transmission $S_{21}$ of resonator R2 when Qubit 2 is prepared in $\gstate$, $\estate$, $\fstate$, and $\h$, respectively. The colored dots represent the measured data while the solid lines show the fitted function (see Supplementary Information I). The vertical lines represent the optimal readout frequencies for the primary (dashed line) and secondary tone (dash-dotted line line). \textbf{(b)} Estimated readout response of  the primary and secondary tones at their respective optimal frequencies. The solid lines represent the in-phase and quadrature data shown in (a). The disks of respective color represent the estimated Gaussian envelope of the signal taking into account the added thermal noise.}
        \label{fig_S21}
\end{figure}

 
With the state-preparation pulses calibrated, we acquire the responses of the readout resonator when the qubit is prepared in $\gstate$, $\estate$, $\fstate$, and $\h$, respectively, illustrated in \figpanel{fig_S21}{a}. We plot the in-phase and quadrature parts of the spectroscopy result, as shown in \figpanel{fig_S21}{b}. We calculate the distance between each point of a pair of trajectories, which represents the separation of the two respective state responses as a function of readout frequency. We find the readout frequencies that maximize $\gstate$-$\estate$ separation for the primary tone, and $\estate$-$\fstate$ separation for the secondary tone. As the readout tones are multiplexed into a single pulse, the frequency and phase of the secondary tone is fine-tuned to minimize the effect on the measurement performance of the primary tone. The amplitudes of both tones are adjusted together to avoid significant readout-induced mixing.\par

\noindent

\subsection{Feedforward neural network}

We utilize a feedforward neural network (FNN) with two hidden layers to discriminate the qubit state using the combined two-tone results. The choice of using an FNN over other methods such as support-vector machine (SVM) or non-linear support-vector machine (NSVM) is justified by the fact that an FNN could achieve greater performance when discriminating more than two states as well as better scalability \cite{Lienhard2022}. An SVM or NSVM will also need to be retrained from scratch every time. While on the other hand, the FNN is capable of transfer learning, where retraining of the network during future re-calibration of the system is significantly more efficient \cite{Bengio2012}. The advantage of using FNN is significant enough to warrant a wider application \cite{Duan2021, Navarathna2021, Lienhard2022}.

The network is implemented with Wolfram Mathematica. The input layer contains four nodes corresponding to the in-phase and quadrature components of the two-tone results, $\{I_{1}[n],\:Q_{1}[n],\:I_{2}[n],\:Q_{2}[n]\}$, of each individual single shot measurement $n$. The first hidden layer contains 16 nodes, while the second layer has 8 nodes. Each node consists of the hidden layers is filtered by a scaled exponential linear unit
(SELU), which acts as the nonlinear activation function. Finally, the output layer contains three nodes, representing the probability of the qubit being in state $\gstate$, $\estate$, and $\fstate$, respectively. We specify an epoch of 100 and learning rate of 0.0005 with a batch size of 64 as a starting point. The network is then trained with 8000 samples and validated by 2000 samples. 

\section*{Data availability}

All relevant data and figures supporting the main conclusions of the document are available on request. Please refer to Liangyu Chen at liangyuc@chalmers.se.

\section*{Acknowledgements}
We would like to thank Daryoush Shiri for valuable discussions on microwave theory and simulation. The fabrication of our quantum processor was performed in part at Myfab Chalmers and flip-chip integration was done at VTT. We are grateful to the Quantum Device Lab at ETH Z\"{u}rich for sharing their designs of printed circuit board and sample holder. The device simulations were enabled by resources provided by the Swedish National Infrastructure for Computing (SNIC) at National Supercomputer Centre (NSC) partially funded by the Swedish Research Council through grant agreement no. 2018-05973. This research was financially supported by the Knut and Alice Wallenberg Foundation through the Wallenberg Center for Quantum Technology (WACQT), the Swedish Research Council, and the EU Flagship on Quantum Technology H2020-FETFLAG-2018-03 project 820363 OpenSuperQ.\par

\section*{Author Contributions}

L.C. performed the measurements and analysis of the results; L.C. and H.L. designed and simulated the device; L.C. and Y.L. developed the idea and carried out theoretical simulations of the system. L.C., C.W.W., and C.J.K. contributed to the implementation of the control methods. S.A. and B.L. developed the machine learning algorithm. S.K., M.R., A.O., J.Bi., A.F.R., M.C., K.G., L.G., and J.G. participated in the fabrication of the device. L.C., G.T. wrote and A.F.K., P.D., and J.By. edited the manuscript. A.F.K., P.D., J.By., and G.T. provided supervision and guidance during the project. All authors contributed to the discussions and interpretations of the results.\par

\section*{COMPETING INTERESTS}
The authors declare no competing interests.\par


\bibliography{library}
\bibliographystyle{revtex.bst}

\end{document}